\documentclass[review]{elsarticle}

\usepackage{lineno,hyperref}
\modulolinenumbers[5]
\usepackage{color}
\usepackage{ulem}
\usepackage{rotating}
\usepackage{tabularx,ragged2e,booktabs,caption,graphicx,epstopdf}
\newcommand{\gao}{$\beta$-Ga$_2$O$_3$}
\newcommand{\sio}{SiO$_2$}
\newcommand{\alo}{Al$_2$O$_3$}
\newcommand{\z}{$\zeta$}

\journal{Carbon}
\begin{document}
\begin{frontmatter}
\author[1]{Soumen Mandal}
\ead{mandals2@cardiff.ac.uk}
\author[2]{Karsten Arts}
\author[2,3]{Harm C. M. Knoops}
\author[1]{Jerome Cuenca}
\author[1]{Georgina Klemencic}
\author[1]{Oliver A. Williams}
\ead{williamso@cardiff.ac.uk}
\address[1]{School of Physics and Astronomy, Cardiff University, Cardiff, UK}
\address[2]{Eindhoven University of Technology, 5612 AZ Eindhoven, Netherlands}
\address[3]{Oxford Instruments Plasma Technology, North End, Yatton, Bristol, BS49 4AP, UK}
\title{Surface zeta potential and diamond growth on gallium oxide single crystal}

\begin{abstract}
In  this work a strategy to  grow diamond on \gao{}  has been  presented. The $\zeta$-potential of the \gao{} substrate  was measured and it was found to be negative  with an isoelectric point at  pH $\sim$ 4.6. The substrates were  seeded with mono-dispersed diamond solution for  growth of diamond. The seeded substrates were etched when  exposed to diamond  growth plasma and globules of gallium could be seen on the surface. To  overcome the problem  $\sim$100 nm of \sio{} and \alo{} were deposited using atomic layer deposition.  The  nanodiamond seeded \sio{} layer was effective in protecting   the \gao{}  substrate and thin  diamond layers could be grown. In  contrast \alo{} layers were damaged when  exposed to  diamond growth plasma. The thin diamond layers were characterised with scanning electron microscopy  and Raman spectroscopy. Raman spectroscopy  revealed the diamond layer to be under  compressive stress of 1.3 -- 2.8GPa.
\end{abstract}


\end{frontmatter}

\section{Introduction}
	In the semiconducting industry silicon is the most widely used material. The band gap of Si is well suited for tailoring the conductivity from semi-insulating to conducting. It also allows for the formation of both n-type and p-type material. As a result the application landscape of Si in the semiconducting industry is vast. Nonetheless, there are areas of applications in which Si as a material is not well suited. For example, many high temperature applications are dependent on the breakdown electric field strength (E$_{br}$), which has a strong correlation with the bandgap. To overcome the limitations of silicon in high temperature applications, many wide bandgap (2.0 eV $\leq E_g \leq$ 7.0eV)\cite{casa98} compound semiconductors like SiC (E$_g = 2.4-3.2$ eV)\cite{casa98} and GaN(E$_g = 3.39$ eV)\cite{mish08} have been developed. Even as wide band gap electronics based on SiC and GaN are maturing, newer materials with wider band gaps (E$_g > 3.4$ eV) are appearing on the horizon. One of the ways to compare different semiconductors is through figures of merit. For low frequency operations,  the Baliga figure of merit(BFM)\cite{bali82} is widely used and for high frequency operations Johnsons figure of merit(JFM)\cite{john65} is widely accepted. Based on BFM of different wide band gap materials, \gao{}, AlN, diamond and c-BN are some of the materials that are considered superior to GaN and SiC\cite{tsao18}. Of these four materials, \gao{} has attracted a lot of attention in recent times\cite{tsao18, higa16, pear18, pear18a, higa18, xue18}, mostly due to availability of large substrates. These large substrates are grown by several melt growth methods like float zone\cite{vill04, ohir06}, Czochralski\cite{tomm00, gala17}, Bridgman\cite{hosh16, ohba16} and edge-deﬁned ﬁlm-fed growth (EFG)\cite{higa16, aida08, kura16} .
	
	While the study in growth of large substrates is an on-going topic, researchers have already demonstrated devices made from gallium oxide\cite{higa12, higa13, chab16, gree16, wong16, mose17, zhou17, noh19, kim20}. The demonstrated devices point towards promising advantages in gallium oxide over traditional wide band gap semiconductors like SiC and GaN. However, as with most high power devices, thermal management in gallium oxide devices, with its low thermal conductivity (10-30  W/mK)\cite{guo15}, is a major bottle neck in the development of technology based on this material\cite{higa18}. It is a well known fact that higher operation temperature in high power devices can lead to significantly lower lifetimes\cite{lee08, pom15}.  One way researchers have overcome this problem is by exfoliating gallium oxide\cite{noh19, kim20, chen19} or by growing gallium oxide on single crystal diamond(thermal  conductivity $\sim$ 2000 W/mK) substrates\cite{chen20}. Another approach that has been recently demonstrated is low temperature bonding of single crystal diamond on \gao{} \cite{mats20}. The major drawback in all these approaches is the small area of the resulting heterostructure for device fabrication. Availability of large area single crystal is currently very limited\cite{schr17} or very expensive. An alternative approach can be direct bonding of large \gao{} crystals to chemical mechanical polished\cite{thom14, werr17, mand18} flat polycrystalline diamond films. This technique is also not straightforward and will involve non-trivial sample surface preparations. A solution to the thermal management problem can be to grow a thick diamond layer on \gao{} single crystal similar to what we have shown in the past on AlN thin films\cite{mand19}.	
		
	The growth of diamond on non-diamond substrates is not trivial. The surface energy of diamond is $\sim 6$ J/m$^2$\cite{hark42}. In comparison the surface energy of silicon, which is most commonly used substrate, is $\sim 1.5$ J/m$^2$ \cite{jacc63}. Due to this large surface energy difference heteroepitaxial growth of diamond is not possible\cite{will11rev, mandrev} and results in only isolated diamond islands. On silicon the density of such islands are of the order of $10^4-10^5$cm$^{-2}$. With a surface energy of $\sim 1$ J/m$^2$ \cite{berm06} for \gao, similar island formations are expected for heteroepitaxial growth. Hence, a seeding/nucleation technique is essential for growth of diamond films on gallium oxide surfaces. An additional issue with growth of diamond on non-diamond substrates using typical CVD approaches is the exposure of the sample to chemically reactive H$_2$ plasma. For example, there are well-known challenges for depositing diamond on GaN\cite{mand19}. Additionally, H$_2$ plasmas are also known to cause surface damage to \gao\cite{poly19}. In this work,  growth of diamond on \gao{} substrates has been demonstrated. The $\zeta$ - potential of the substrates have been measured to determine the best seeding solution. It was found that the substrates disintegrate when exposed to plasma during diamond growth leaving behind globules of gallium. To overcome this, a thin layer of \sio{} and \alo{} was deposited as a buffer layer. The $\zeta$-potential of the buffer layers were also measured. Thin diamond films were grown (250nm) and characterised using scanning electron microscopy. Also, the film stresses were characterised using Raman spectroscopy. For deposition of thick diamond layers alternative strategies will have to be designed  to manage the stress at the interface, which will be the basis of further studies.


\section{Experiment}
When a solid is immersed in a liquid it  acquires a surface charge which  is compensated by  ions of opposite charge or counterions  that are loosely  bound to  the surface. The surface charge is formed due to  adsorbed ions on the surface of  solid and that forms the first  layer. The  counterions are then  attracted towards the surface due to Coulomb attraction and are part  of  a second layer called diffuse layer. The first layer  is immobile with respect  to solid. The surface charges  also generate  a potential which  decays with distance from the solid surface. $\zeta$-potential is defined as the potential at  the boundary of the immobile and mobile liquid with  respect to rest of the  liquid. For measuring $\zeta$-potential  it is essential to measure the charge  on the counterions. If a channel  is formed between two  solid surfaces and an electrolyte is passed through the channel, the counterions experience a shearing force and move along with  the electrolyte\cite{wag80}. As a result a charge separation is formed between the inlet and outlet of the channel. This leads to formation  of an electric potential known  as streaming  potential. First  suggested  by Van  Wagenen et al.\cite{wag80}, it has been  used for determination of $\zeta$-potential of variety of surfaces\cite{mand19, voi83, nor90, sca90, mand17, bla19, mand19a, blan21}. The  Helmholtz-Smoluchowski equation  gives a relation between $\zeta$-potential and streaming current/potential\cite{wern98}. By measuring the change in  streaming current/potential as a function  of electrolyte  pressure the $\zeta$-potential can be determined. The $\zeta$-potential of \gao{}, \sio{} and \alo{} coated \gao{}, sapphire and quartz were measured using Surpass$^{TM}$ 3 electrokinetic analyzer. In the present work the channel width was kept between 90-110 $\mu$m. The electrolyte was 10$^{-3}$ M solution of KCl in DI water with electrolyte pressure varying between 200 and 600 mbar. The pH  of the electrolyte was varied by  addition of 0.1M NaOH and 0.1M HCl solution with the inbuilt titrator in Surpass$^{TM}$ 3. 

The $\sim$100 nm \sio{} and \alo{} layers were grown by atomic layer deposition (ALD), using an Oxford Instruments FlexAL reactor\cite{heil07}. Both layers were grown using 1100 ALD cycles and a table temperature setpoint of 300 $^o$C. The \sio{} layer was grown using SiH$_2$(NEt$_2$)$_2$ (bis(diethylamino)silane) as precursor and O$_2$ plasma as coreactant, where the O$_2$ plasma was generated by an inductively coupled plasma source. Purely thermally-driven ALD was employed for the growth of \alo, using Al(CH$_3$)$_3$ (trimethylaluminum) and H$_2$O. The thicknesses of the deposited \sio{} and \alo{} layers were measured by spectroscopic ellipsometry (SE), using a M–2000D Spectroscopic Ellipsometer of J.A. Woollam Co. The SE measurements were performed ex-situ on silicon substrates (Czochralski silicon (100) with $\sim$1.6 nm native oxide) that were processed alongside the \gao{} substrates. A Cauchy dispersion model was used for fitting the SE data\cite{lang09}. The film thicknesses were measured to be 89 nm for the \sio{} layer and 111 nm for the \alo{} layer. Moreover, the refractive indices of the films (at 632.8 nm) were measured as 1.45 for \sio{} and 1.65 for \alo.

After $\zeta$-potential measurements the substrates  were seeded with mono-dispersed  diamond solution. The details of the seeding process can be found here\cite{mandrev}.  The  seeded wafers were then introduced in a Carat Systems SDS6U microwave plasma chemical vapour deposition(MPCVD) systems. The growth  was done in  two steps, that is incubation and growth. The growth of diamond by MPCVD is essentially a complex process of growth and etching  of all forms of carbon with  non-diamond carbon surviving only in negligible quantities. The  nanodiamond seeds used for seeding are typically  5nm in size. As a result  they  have considerable surface to volume ratio resulting in large proportion of non-diamond carbon\cite{will11rev} at the  surface which  are easily etched in the plasma. The incubation  step, lasting for 3-10 mins, is used for fast growth of seeds so that they are not  fully etched away during initial stages of growth. For the incubation step a gas mixture of 5\% CH$_4$/H$_2$ and 50 torr chamber pressure with 3.7 kW  microwave power was used. The  incubation period was 5 minutes and the gas flow was maintained at  500  sccm. After the incubation period the gas mixture was modified to 3\% CH$_4$/H$_2$. The total time for the growth was 40 mins and the growth temperature was 630 $^o$C as recorded  by  dual wavelength  Williamson  pyrometer. After growth  the sample was slowly  cooled in hydrogen plasma.  The substrates, diamond layers and oxide coating were analysed using a Horiba LabRAM HR Evolution equipped with SynapsePlus Back-Illuminated Deep Depletion (BIDD) CCD. The spectroscope is equipped with three lasers having wavelengths of 473, 532  and 660nm. Data  from all three lasers have been included in this work. A Hitachi SU8200 series scanning electron microscope (SEM) operating at 10 kV and working distances between 9 and 11 mm was used for imaging the samples after  growth.  ($\bar2$01) \gao{} substrates  used in this work were commercially  sourced from Tamura Corporation. The substrates  were single side chemical mechanical polished. 10X10mm and 5X5mm pieces were cut from a 2" wafer  for growth of diamond.


\section{Results and Discussion}

\subsection{Zeta potential measurement}
\begin{figure}[ht]
\centering
\includegraphics[width = 7cm]{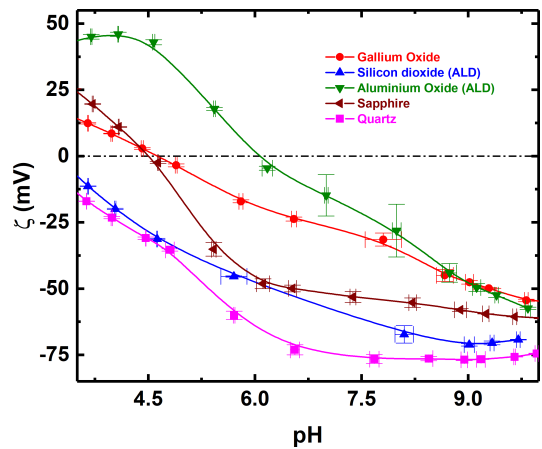}
\caption{Zeta potential vs pH for \gao{} crystal. The figure also show the zeta potential of deposited buffer layers namely, \sio{} and \alo{}. The zeta potential of quartz and sapphire are shown for comparison as well. The solid lines are a guide to the eye. The dashed black line has been put to mark the isoelectric point.} \label{zeta}
\end{figure}

The determination of the $\zeta$-potential of a non-diamond substrate, which is directly related to surface charge, is essential for the determination of the type of seed solution needed for high seed density. It is possible to create diamond seed solution with positive or negative surface charge in water\cite{will11rev, mandrev, hee11}. For high seed density seed solution with charged particles opposite to charges on substrate surface are essential. Figure \ref{zeta} shows the $\zeta$-potential as a function of electrolyte pH for \gao{} along with potentials of ALD deposited \sio{}, \alo{}, quartz and sapphire. The $\zeta$-potential of \gao{} is negative beyond the isoelectric point around pH $\sim$ 4.6. This is in contrast to gallium oxide nanoparticles which have an isoelectric point of pH$\sim$9 \cite{kosm01}. However, $\zeta$-potential of single crystal \gao{} plate is similar to the $\zeta$-potential for Ga faced GaN\cite{mand17} where as, N-face GaN has more negative $\zeta$-potential. The higher negative $\zeta$ of N-faced GaN could be down to higher amounts of adsorbed oxygen on the surface.  However, the main point of interest for seeding is around pH 6-7. In this range the $\zeta$ potential of \gao{} is negative with a value between -20 and -30mV. This means that for high seed density H-terminated diamond seed solution is needed\cite{hee11}. The substrates were seeded by dipping in H-terminated nanodiamond solution. The seeded substrates were then exposed to H$_2$/CH$_4$ plasma for diamond growth. As soon as the substrates were exposed to the plasma, immediate damage to the \gao{} surface could be observed. On closer examination under SEM (see later in figure \ref{sem}), globules of gallium metal was observed. 

To overcome the damage to the substrate surface from the plasma, it was encapsulated with a dielectric coating. For the purpose of the current study, \sio{} and \alo{} were chosen. Thin layers of the coatings were deposited with ALD. For measuring the zeta potential of the deposited coatings, quartz pieces were included during the deposition of the dielectrics on \gao{}. The $\zeta$ potential of 100nm ALD deposited dielectrics are shown in figure \ref{zeta}. For comparison $\zeta$-potential of the quartz  (\sio{}) and sapphire  (\alo{}) are also shown.  While \sio{} has negative \z-potential over the whole measurement range (pH $\sim$ 3-10), \alo{} has isoelectric point at pH $\sim$ 6 and has negative \z-potential beyond that point. In the case of \sio{} the \z-potential is not dependent on whether it is in the form of quartz, thin layer or nanoparticles\cite{chen17}. In contrast, for \alo{} it has a strong dependence on whether it is in the form of sapphire (isoelectric point pH $\sim$ 4.5), thin film (isoelectric point pH $\sim$ 6) or nanoparticles (isoelectric point pH $\sim$ 9)\cite{bahe02}, which is similar to what is seen for gallium oxide. As before the substrates with dielectric encapsulation were seeded with H-terminated diamond seed solution. The seeded samples were exposed to diamond growth conditions. Thin diamond films could be grown on substrates encapsulated with \sio{}. The \alo{} thin layer disintegrated on exposure to plasma and exposed the underlying \gao{} surface. This is probably due to large thermal mismatch between \alo{} layer and \gao{}. Alternatively, it is  also  possible that  \alo{} is easily  etched\cite{lee98} in CH$_4$/H$_2$ plasma while \sio{} can  act as a mask\cite{lee93} in the same environment.
\subsection{Scanning Electron Microscopy}
\begin{figure}
\centering
\includegraphics[width = 7cm]{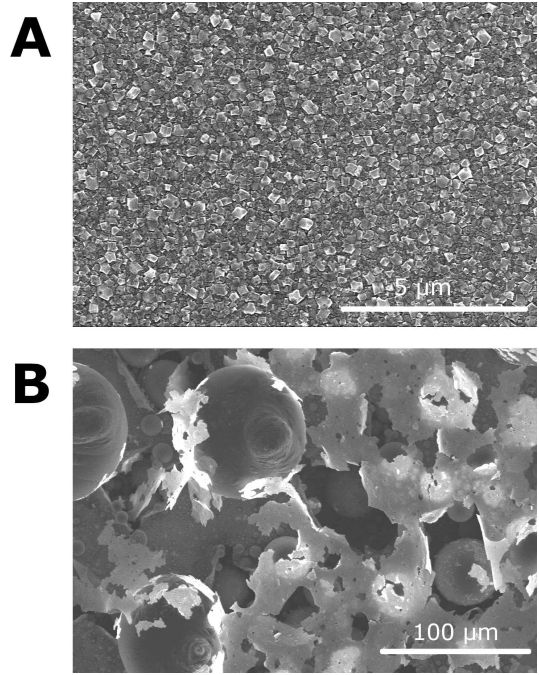}
\caption{A. Scanning electron microscope image of thin diamond film grown on \sio{} coated \gao{}. B. Image of the \alo{} coated \gao{}  surface after exposure to diamond growth condition. Tiny  gallium globules along with flakes of \alo{} can be clearly seen.} \label{sem} 
\end{figure}
Figure \ref{sem}A shows the scanning electron microscope images for diamond thin films grown on \sio{} coated \gao{}. A fully  coalesced film with grain sizes of the order of 250nm can be clearly  seen. Panel B in the image shows sample surface coated with \alo{}. The \alo{} coating has completely disintegrated in the diamond growth plasma. Small flakes of \alo{} can be clearly seen in the image. Once the \alo{} layer disintegrates, the \gao{}  surface is exposed to plasma and small gallium globules are formed. Such globules were also seen for samples that were not coated with a protective oxide.   

\subsection{Raman spectroscopy}

\begin{figure}
\centering
\includegraphics[width = 7cm]{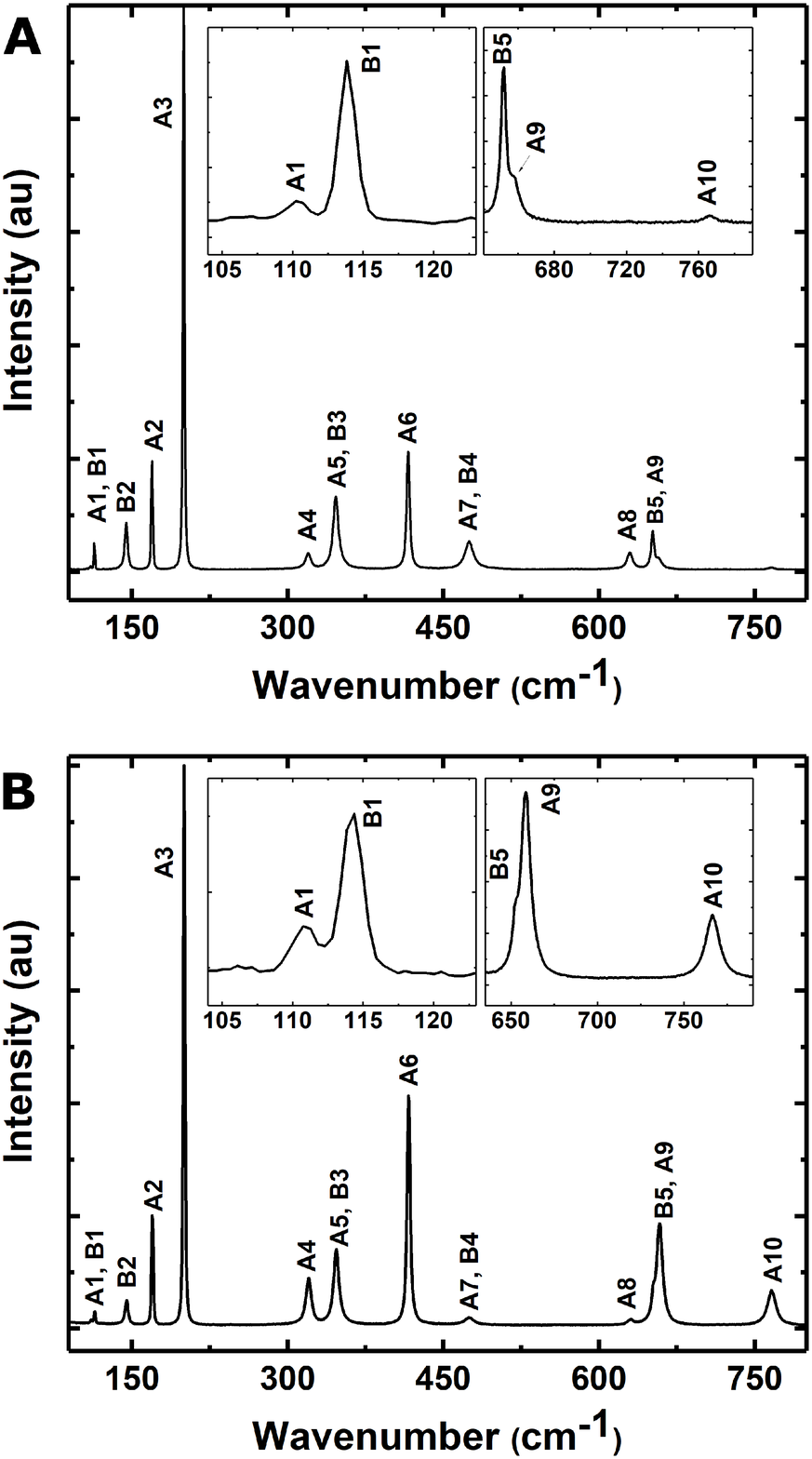}
\caption{Raman spectroscopy data  is shown for A) \gao{}  and B) \sio{} covered $\beta$-Ga$_2$O$_3$. The common peaks are marked in the figure. The inset shows the magnified view of A1, B1 peaks (left inset)  and B5, A9 and A10 peaks (right inset). Note the change in relative intensities of various peaks in bare  substrate  and \sio{} covered substrate, specifically  A6, B5, A9 and A10. The data was taken using 532 nm laser.} \label{ramg} 
\end{figure}

In this section we will discuss the Raman  spectra of \gao, \sio{} coated \gao{} and diamond grown on \sio{} coated \gao. A  detailed comparison between \gao{} and \sio{} coated \gao{}  will be done to investigate the effects of the \sio{} layer on the Raman modes of \gao. \gao{} unit cell has 30 phonon modes \cite{dohy82, kran16}, of these only 27 are optically active. Amongst the optically active modes 15 are Raman active which have A$_g$ and B$_g$ symmetry. Since only the modes with A$_g$ and B$_g$ symmetry are Raman active in \gao{}, A$_g^{(1)}$ will be denoted simply as A1, where A$_g^{(1)}$ has a spectral position at 111 cm$^{-1}$ according to Kranert et al.\cite{kran16}. The rest of the Raman modes will follow similar convention throughout this article. The Raman spectra for bare \gao{}  substrate and 100nm \sio{} covered \gao{} is shown in  figure \ref{ramg}A and B respectively. The  data was taken  using 532 nm laser.  The data taken with  473 and 660nm  lasers  are  presented in figures  S1 and S2 respectively. The common peaks in \gao{}  are marked in  the figure. The peak positions of the different modes for both samples are listed in table \ref{tab1} along with their full width at half maximum (FWHM). The peak positions have been determined by fitting a Lorentz function to the Raman data\cite{kran16}. While the bare substrate data fits well with the Lorentzian peak fit, some peaks in \sio{} covered substrate are best fitted with Voigt function.  The results of the peak fitting using Voigt function are shown in table S1. The FWHM of the peaks fitted with Voigt function have been calculated on the basis of definition given by Olivero et al.\cite{oliv77}. GLmix in the table (Table S1) defines the proportion of Lorentzian character of the Voigt function\cite{sher19}.  The relative intensities (Rel. Int.) of the peaks are calculated from the area of the fitted curves and has been normalised with respect to A3 peak. The peaks of B3 and B4 modes are very close to A5 and A7 respectively and is indistinguishable from each other. Hence, the modes A5, B3 and A7, B4 are represented together in the table.  The relative intensities on \gao{} show trends similar to that shown by Kranert et al.\cite{kran16} In contrast, there is considerable deviation in the intensity trends for \sio{} covered \gao{} and follows more closely the intensity trend of parallel polarisation along the [102] direction on  ($\overline201$) \gao{}.

\begin{sidewaystable}
\centering
\begin{tabular}{ c |c c c| c c c}
\toprule 
Raman mode & \multicolumn{3}{c|}{\gao{}} & \multicolumn{3}{c}{\sio{} on \gao{}} \\ 
 \midrule
 &Peak  Center & FWHM & Rel. Int. & Peak  Center & FWHM & Rel. Int. \\
A1&109.95 &2.30&8.70&110.67&2.27&7.66\\  
B1&113.89&1.31&35.29&114.24&1.68 &19.67\\
B2&144.54&3.15&144.97&144.93&3.39 &72.91\\  
A2&169.44&1.67&189.92&169.84&1.84&195.13\\  
A3&199.99&1.70&1000&200.41&1.90&1000 \\
A4&319.85&5.69&83.19&320.36&5.94&241.01 \\  
A5, B3&346.56&6.03&419.67&347.00&6.06&398.24\\
A6&416.28&3.27&378.74&416.75&3.39&701.80 \\
A7, B4&475.03&9.76&256.77&475.15&10.28&62.19 \\  
A8&629.77&6.46&101.68&630.59&7.02& 28.81 \\  
B5&651.88&3.49&121.95&652.34&3.47& 54.73\\  
A9&658.04&6.61&51.97&658.50&6.78&604.38\\  
A10&766.20&8.43&13.70&766.53&9.69 &295.99\\  
\bottomrule

\end{tabular}
\caption{Peak positions, FWHM and relative intensity with respect to A3 of Raman peaks for \gao{} and \sio{} coated \gao{}. The peak positions and FWHM are in cm$^{-1}$.}\label{tab1}
\end{sidewaystable}

\begin{figure}
\centering
\includegraphics[width = 9cm]{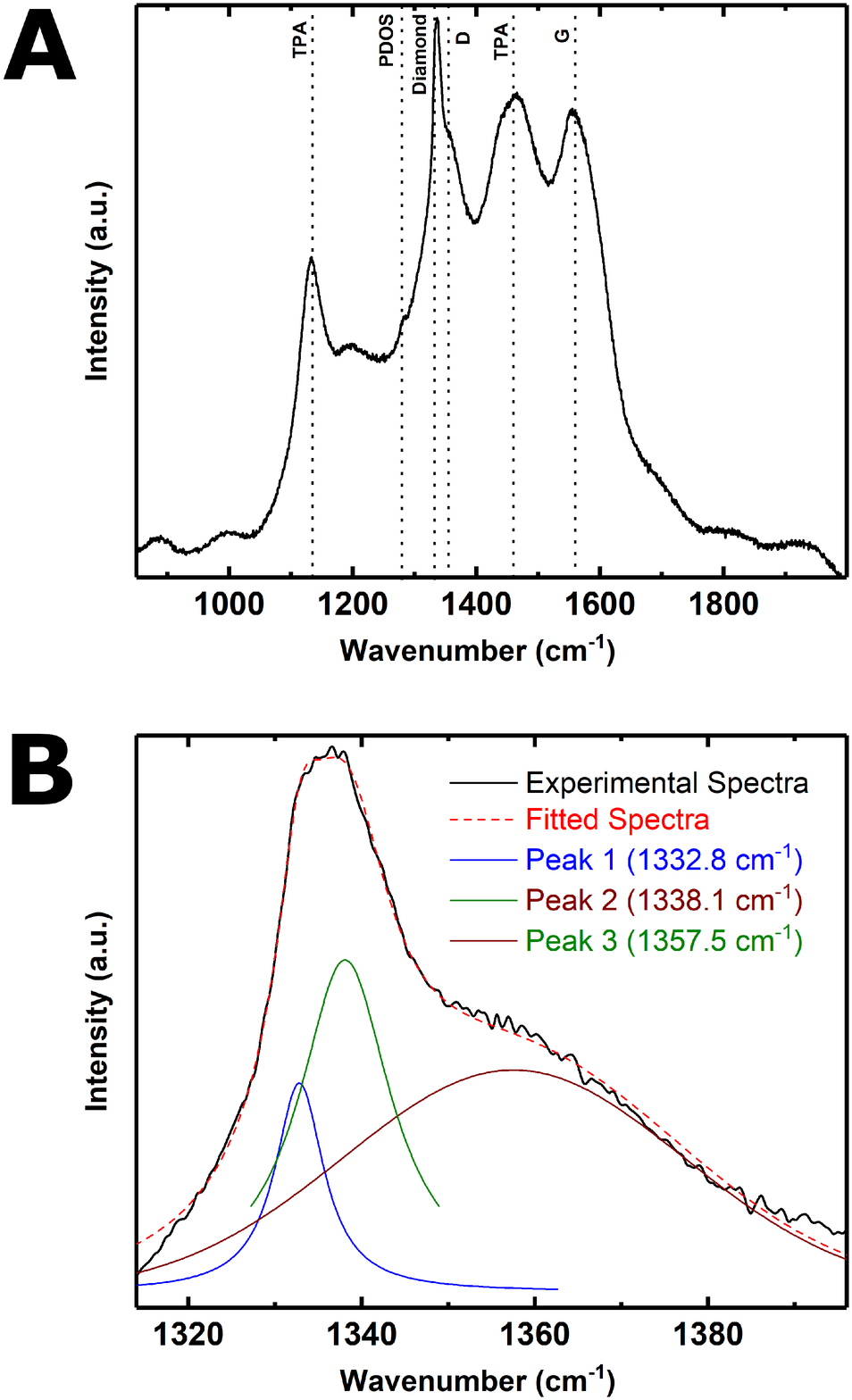}
\caption{A. Raman spectroscopy data  is shown for diamond thin film grown on \sio{} coated \gao{}. The positions of common Raman peaks seen in diamond thin films are marked in the figure. The data was taken using 532 nm laser. B. The zoomed in view of the diamond peak is shown  along with the fitted curves. The peak has been  fitted with two Lorentzian and one Gaussian curve.} \label{rd} 
\end{figure}

The peaks in the data can be divided into three distinct regions\cite{dohy82}. First region are the peaks below 200 cm$^{-1}$, the second region are the peaks between 300 and 500 cm$^{-1}$ and the third one are the peaks above 600 cm$^{-1}$.  The peaks below 200 cm$^{-1}$ are associated with vibrations of the tetrahedra chains\cite{dohy82} i.e. small amplitudes and have narrow widths of $\sim$3 cm$^{-1}$. In this region the FWHM of the peaks are slightly wider for \sio{} covered \gao{} and this trend is seen for most peaks. The intensity trend is similar except for the B1 and B2 modes having lower intensities. However, the intensity trend for the peaks beyond 200 cm$^{-1}$ is completely different for \gao{} and \sio{} covered \gao{}. In this region the peaks are broader (FWHM $>$ 3 cm$^{-1}$). The modes above 200 cm$^{-1}$ correspond to bending and stretching modes or internal vibrations of the tetrahedra groups. The presence of the \sio{} layer on top of \gao{} seem to have disproportionate effect on these modes. Finally, Raman spectra of diamond film grown on \sio{} covered \gao{} has been taken and the data is presented in figure \ref{rd}. The data  was taken  with 532 nm laser. The  spectra  from \gao{},  \sio{} coated \gao{}  and diamond taken with 532nm laser are presented together  in figure S3 for comparison.

The Raman spectroscopy data for diamond thin film on \sio{} coated \gao{}, shown in figure \ref{rd}A, shows a clear diamond peak. The data was taken with 532 nm excitation laser and a straight line luminescence background was subtracted\cite{dych16} before presenting the data in the figure. Data was also taken with 473 and 660 nm excitation. The data taken  at  473 nm also showed luminescence background, however the background was absent for the data taken at 660 nm\cite{may07}. The \gao{} Raman peaks were present  in all  three dataset and the same are shown in  figure S4. The figure also shows the background straight line subtracted for the data in figure \ref{rd}.   The commonly observed peaks in thin diamond films are marked in figure  \ref{rd}. The small shoulder around 1350 cm$^{-1}$  is associated with the D peak from amorphous carbon \cite{ferr01, fer04, praw04}. The peak at around 1332 cm$^{-1}$ is the diamond peak\cite{ram30, bhag30}, however, it is  clearly shifted towards higher  wavenumbers and is  heavily  convoluted with  additional bands. This is likely due to stress in the diamond thin film\cite{knig89, ager93, wind95, anas99, dych15}. Accurate determination of the stress magnitude requires appropriate deconvolution of the spectra. Gries et al. \cite{grie07} demonstrated that multiple types were required to resolve the spectra; 1 Gaussian for non-diamond carbon and 3 Lorentzians for diamond peaks in various stress states. In the simplest way, resolving the spectra between 1300 and 1400 cm-1 for one Lorentzian and one Gaussian profiles yield peak positions of 1336 cm-1 and 1359.4 cm-1 for the diamond and D peaks, respectively.  In a similar approach to Gries et  al.\cite{grie07}, the spectra in this work obtained a better fit (higher  R$^2$ and lower $\chi^2$) using 2 Lorentzian peaks to resolve the diamond contributions (1332.8 and 1338.1cm-1) and a Gaussian peak to resolve the D peak (1357.5 cm-1). Further increment in number of peaks results in non-convergence of the fitting parameters and leads to unfeasible results. Figure \ref{rd}B shows the zoomed view of the diamond peak and D peak from amorphous carbon. The deconvoluted peaks along with the calculated curve are shown in the figure.

Considering the fits to the diamond peak in figure \ref{rd}B  the shift in the peak is between 4-6  cm$^{-1}$, depending on whether  one considers  the resolved or unresolved peak. Various researchers have estimated different stress coefficients for Raman  peak shifts. Boppart et al.\cite{bop85} estimated the stress coefficient for Raman line shifts to be around 0.38 GPa/cm$^{-1}$ by  measuring the shifts in Raman line for a diamond crystal under  pressure.  Knight  et al. \cite{knig89} concluded a similar  coefficient of 0.42 GPa/cm$^{-1}$ by measuring stress in diamond films. However, Yoshikawa et al. \cite{yosh89, yosh91} used the value of 2.63 GPa/cm$^{-1}$  for  their  results assuming only hydrostatic stress. Windischmann  et al.\cite{wind95} used a value of 0.39  GPa/cm$^{-1}$, however the postulated that films grown on substrates are under  biaxial stress,  hence the coefficient  is 1.5 times larger  resulting in a coefficient of 0.59 GPa/cm$^{-1}$. These figures were later disputed by  Anastassakis \cite{anas99}. Based on equations given by  Anastasskis \cite{anas99},  the stress in  the diamond films presented in  this work is estimated to be between 1.3 and 2.8 GPa (assuming  bisotropic stress), which is closer to the results of Boppart et al.\cite{bop85} and Knight et  al.\cite{knig89}. The stress along with adhesion strength between  various components of  the stack  consisting of \gao{}, \sio{} and diamond will determine if a film will stick on the substrate. It  should be noted that the stress between layers is thickness dependent\cite{cuen21}. In  the case of diamond on \sio{} covered \gao{} the film thickness is limited to 250nm.

Figure \ref{rd}A also shows peaks at $\sim$1150 and $\sim$1450 cm$^{-1}$. These are associated with the  transpolyacetylene present at grain boundaries and diamond surface\cite{tpa01}. The peak around $\sim$1560 cm$^{-1}$ is associated with G  peak from amorphous carbon\cite{fer04, praw04}. Apart from these a small shoulder can  be seen around $\sim$1280 cm$^{-1}$. This  is associated with phonon density  of states (PDOS)  in diamond\cite{pavo93, fer04}. Small  bumps around 900, 1000 and 1220 cm$^{-1}$ can also be seen which can also be attributed to PDOS of diamond\cite{pavo93}.


\section{Conclusion}
In conclusion, the $\zeta$-potential of \gao{} has been  presented. The $\zeta$-potential was found to be negative  above pH $\sim$ 4.6,  enabling the use  of H-terminated diamond seed solution for seeding of the surface.  The $\zeta$-potential is similar to that of Ga-faced gallium nitride. In contrast gallium oxide nanoparticles have largely positive $\zeta$-potential with isoelectric poit at pH$\sim$9\cite{kosm01} A method for growing diamond on \gao{} crystal  has also been detailed. It was found that direct growth of diamond on \gao{} is not possible with MPCVD. As a result,  thin  layer  of \alo{} and \sio{} were  deposited on \gao{} before diamond deposition.  Upon seeding  with diamond solution,  \sio{} layer  survived the diamond growth condition and resulted in a thin diamond layer. In  contrast the seeded \alo{} layer  of similar  thickness was completely destroyed when  exposed  to diamond growth conditions. Raman measurements were done on \gao{}, coated \gao{} and diamond on coated \gao{}. It was  found that  the diamond layer  is under considerable amount of stress and this is due to the mismatch in thermal expansion coefficient between \gao{} and diamond. Further work is needed to manage the  interfacial stress between  diamond and coated \gao{} for growth of  thick diamond layer  on \gao{}.


\section{Acknowledgment}
This project has been supported by Engineering and Physical Sciences Research Council under programme Grant GaN-DaME (EP/P00945X/1). The metadata for the results presented in this paper can be found here().

\bibliography{ref}

\pagebreak

\begin{center}
\textbf{\large Supporting Information: Surface zeta potential and diamond growth on gallium oxide single crystal}
\end{center}
\setcounter{equation}{0}
\setcounter{figure}{0}
\setcounter{table}{0}
\setcounter{section}{0}
\setcounter{page}{1}
\makeatletter
\renewcommand{\theequation}{S\arabic{equation}}
\renewcommand{\thesection}{S\arabic{section}}
\renewcommand{\thefigure}{S\arabic{figure}}
\renewcommand{\thetable}{S\arabic{table}}
\renewcommand{\bibnumfmt}[1]{[S#1]}
\renewcommand{\citenumfont}[1]{S#1}

\section{Raman Spectroscopy}
Table \ref{tabs1} shows the peak position and FWHM of the Raman peaks as determined by Voigtian peak fitting. GLmix is defined as $\frac{\mbox{FWHM(Lorentz)}}{\mbox{FWHM(Lorentz)+FWHM(Gauss)}}$. This means that a pure Lorentzian profile will give a GLmix of 1 and a pure Gaussian will give the value of 0.
\begin{sidewaystable}
\centering
\begin{tabular}{ c |c c c c|c c c c}
\toprule 
Raman mode & \multicolumn{4}{c|}{\gao} & \multicolumn{4}{c}{\sio on \gao} \\ 
 \midrule
 &Peak  Center & FWHM &GLmix & Rel. Int.& Peak  Center & FWHM & GLmix & Rel. Int.\\
A1&110.29 &1.69&0.56&5.23&110.86&2.21&0.35&6.36 \\  
B1&113.89&1.38&0.16&28.58&114.24&1.76&0.21& 16.49\\
B2&144.54&3.22&0.73&161.81&144.92&3.51&0.65& 80.70\\  
A2&169.45&1.80&0.43&185.33&169.84&1.97&0.42& 191.48\\  
A3&199.99&1.85&0.45&1000&200.41&2.06&0.42 &1000\\
A4&319.84&6.19&0.53&88.01&320.36&6.22&0.66 &272.59\\  
A5, B3&346.40&6.03&1.00&485.66&347.01&6.02&0.87&468.66\\
A6&416.28&3.36&0.71&422.81&416.75&3.50&0.68&792.22 \\
A7, B4&475.03&9.76&0.92&296.33&475.15&10.25&0.75 &69.62\\  
A8&629.70&6.50&1.00&118.22&630.62&6.91&0.92 &33.10\\  
B5&651.88&3.46&1.00&139.91&652.32&3.54&0.78&64.48\\  
A9&657.95&6.87&1.00&62.65&658.50&6.81&0.86 &710.74\\  
A10&766.20&8.43&1.00&15.90&766.53&9.78&0.80 &343.70\\  
\bottomrule

\end{tabular}
\caption{Position of various Raman modes in \gao and \sio covered \gao along with their FWHM and relative intensities with respect to A3 mode. The peak positions and FWHM are in cm$^{-1}$.} \label{tabs1}
\end{sidewaystable}

\begin{figure}[h]
\centering
\includegraphics[width = 7cm]{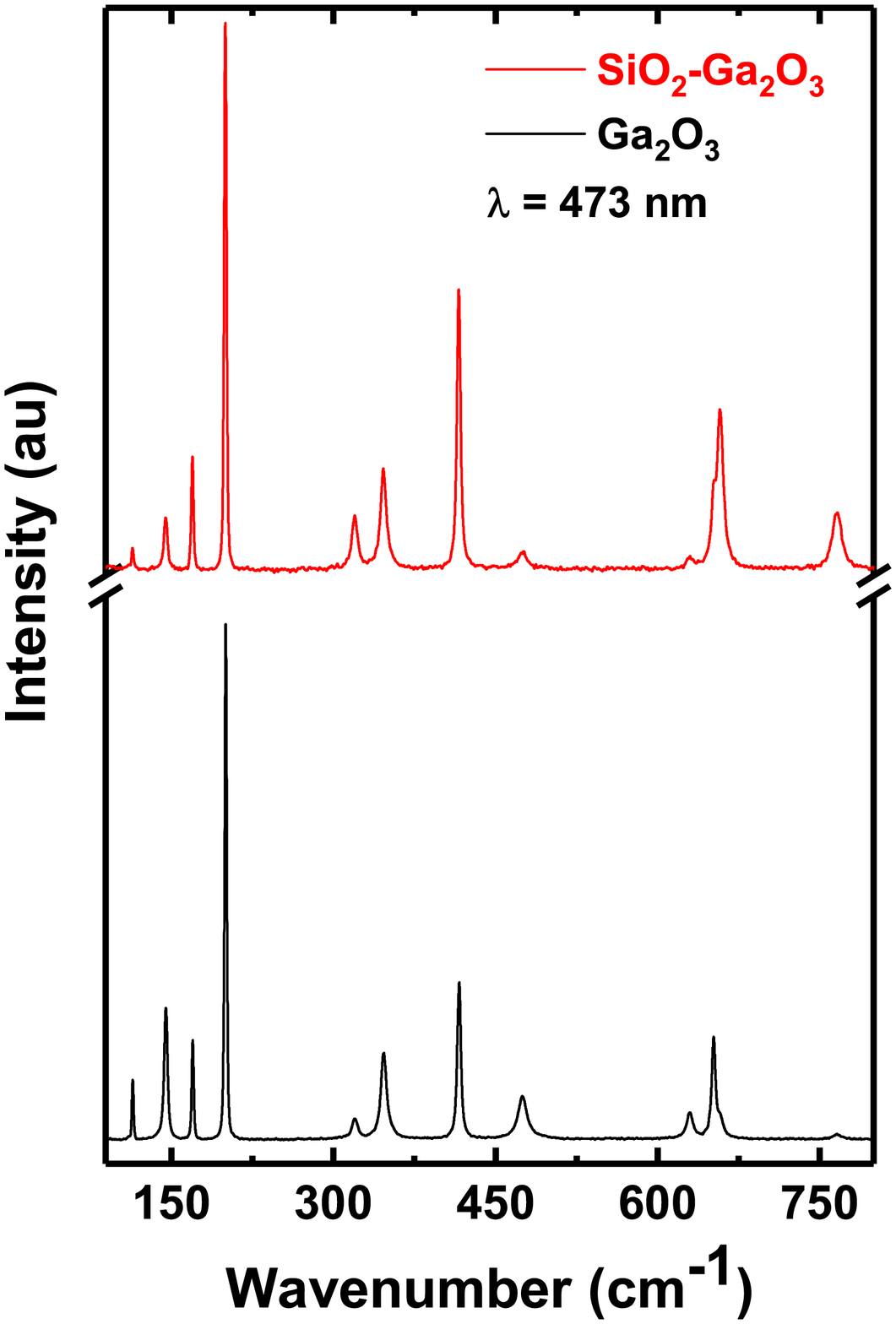}
\caption{Raman spectroscopy data  is shown for \gao(black curve)  and  \sio covered $\beta$-Ga$_2$O$_3$ (red curve) taken with 473 nm laser.} \label{r473} 
\end{figure}

\begin{figure}
\centering
\includegraphics[width = 7cm]{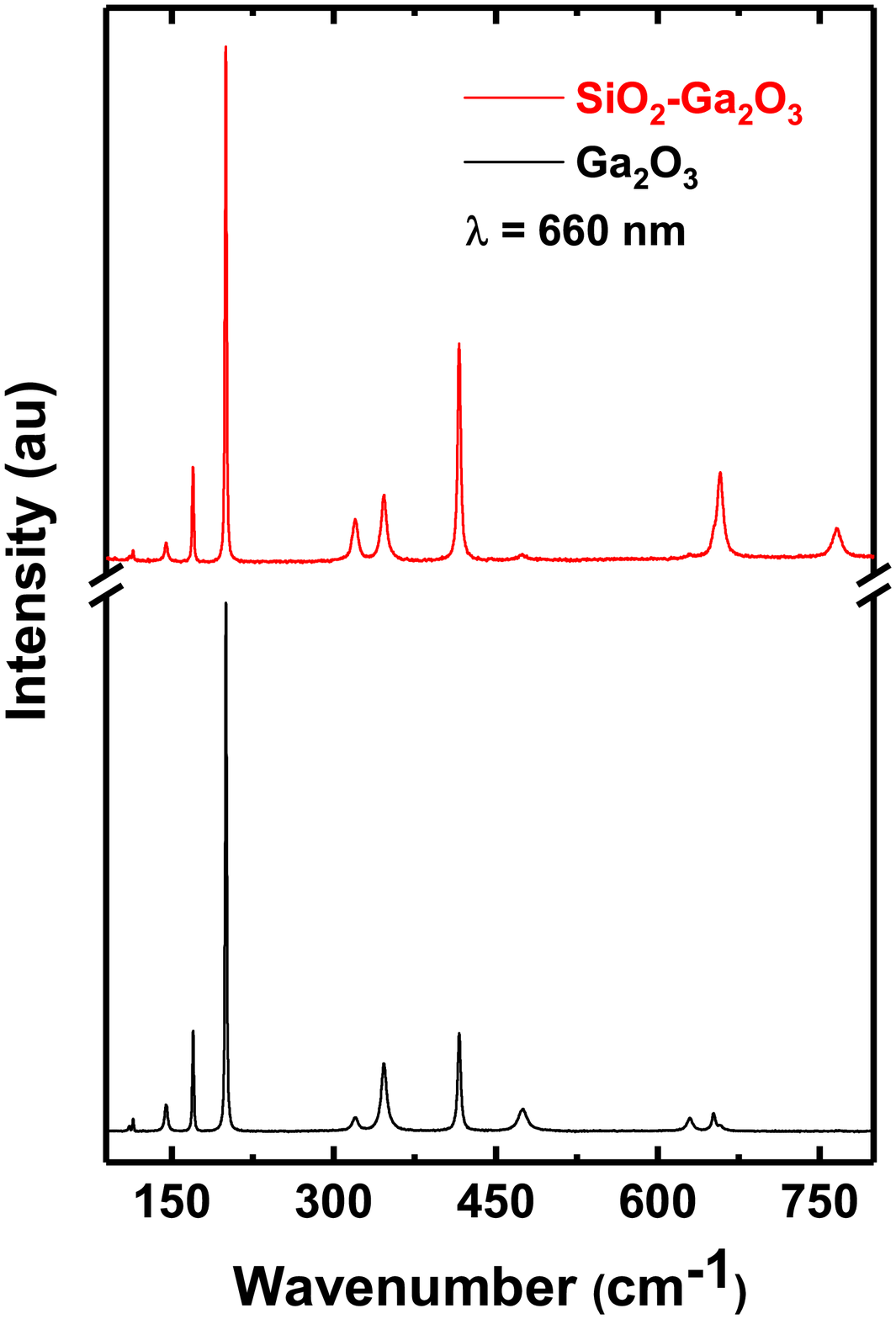}
\caption{Raman spectroscopy data  is shown for \gao(black curve)  and  \sio covered $\beta$-Ga$_2$O$_3$(black curve) taken with 660 nm laser.} \label{r660} 
\end{figure}

Figure \ref{r473} and \ref{r660} shows the Raman spectroscopy data taken with 473nm and 660nm excitation laser respectively. The data on bare \gao is shown in black and the red curves are for data on \sio coated \gao. The change in relative intensities of various peaks are clearly visible between bare substrate and \sio coated substrate.

\begin{figure}
\centering
\includegraphics[width = 7cm]{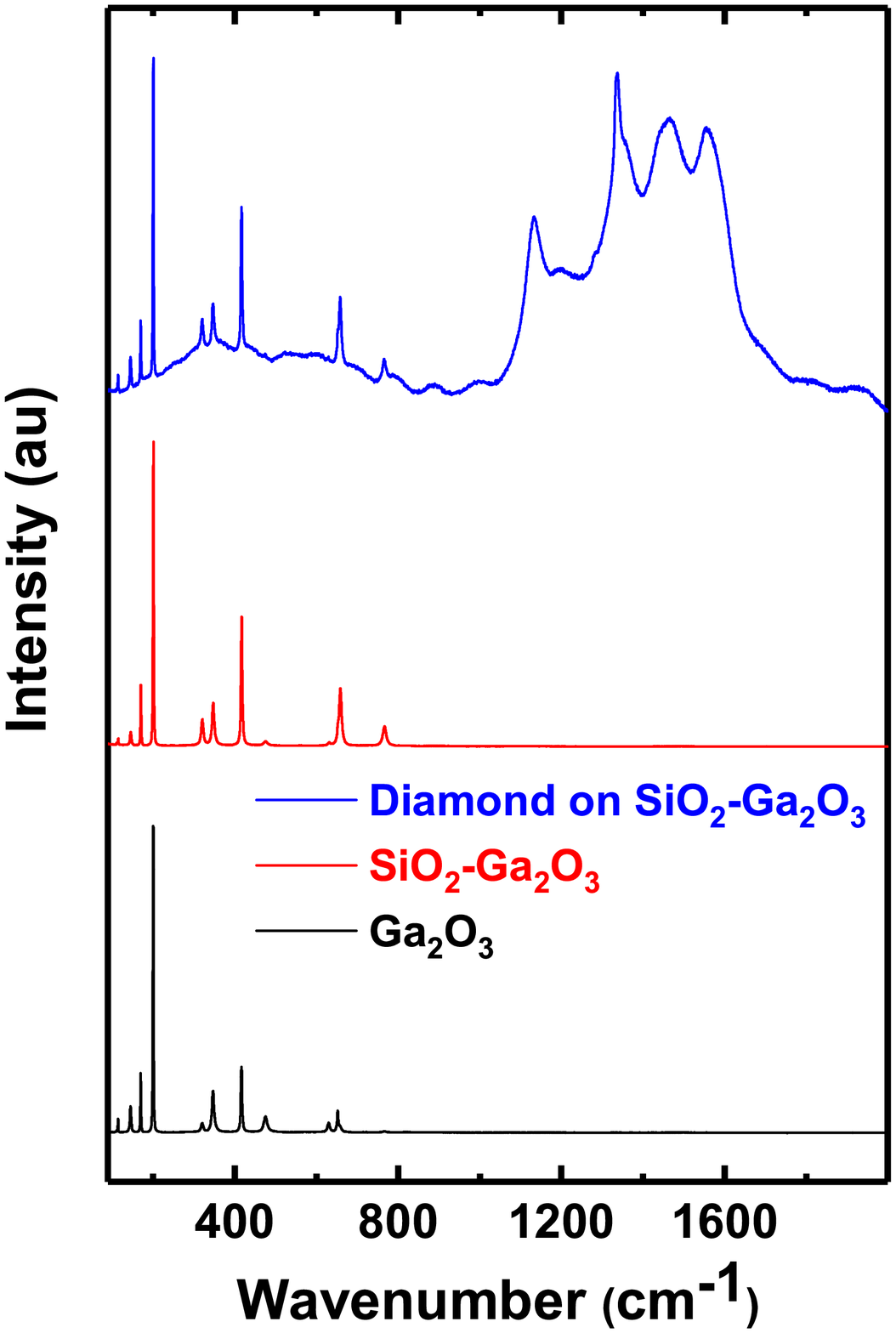}
\caption{Raman spectroscopy data is shown for \gao, \sio covered \gao and diamond grown on \sio covered \gao taken using 532 nm laser. The peaks for \gao are clearly seen along with diamond peaks in the data.} \label{a532} 
\end{figure}

\begin{figure}
\centering
\includegraphics[width = 7cm]{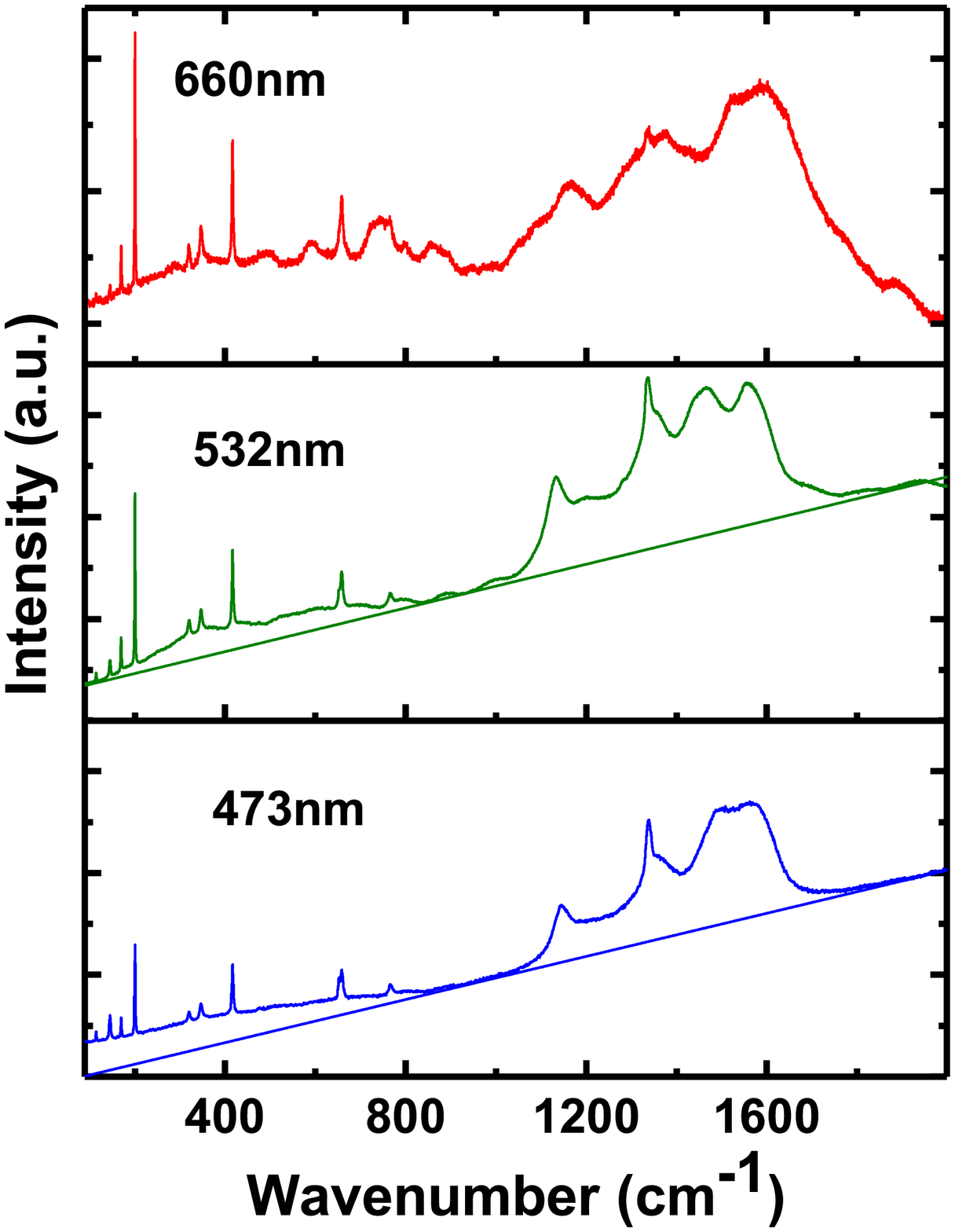}
\caption{Raman spectroscopy data is shown for diamond grown on \sio covered \gao taken using 473, 532 and 660 nm laser. The peaks for \gao are clearly seen along with diamond peaks in the data. A straight line background has been subtracted from the data taken by 473 and 532 nm excitation laser as indicated in the figure.} \label{rdiab} 
\end{figure}

Raman spectroscopy data comparing \gao, \sio coated \gao and diamond grown on \sio coated \gao are shown in figure \ref{a532}. The \gao peaks can be clearly seen all three spectra confirming the \gao layer under the diamond. Figure \ref{rdiab} shows the Raman data of thin diamond film grown on \sio coated \gao. A straight line background has been subtracted from the data taken with 473 and 532nm excitation laser. The data clearly shows the presence of the \gao peaks in the low frequency regime. 

\end{document}